\documentclass[]{pasj01}
\usepackage{comment}

\draft

\begin{document} 
\Received{2017/8/24}%{yyyy/mm/dd}
\Accepted{2018/2/14}%{yyyy/mm/dd}
%\Published{yyyy/mm/dd}

\title{The formation of a Spitzer bubble RCW 79 triggered by a cloud-cloud collision}

%%% begin:list of authors
% Do NOT capitalize all letters in "textsc".
\author{Akio \textsc{Ohama}\altaffilmark{1}$^{*}$}
\altaffiltext{1}{Department of Physics, Nagoya University, Furo-cho, Chikusa-ku, Nagoya, Aichi 464-8601, Japan}
\altaffiltext{2}{Nobeyama Radio Observatory, National Astronomical Observatory of Japan (NAOJ), National Institutes of Natural Sciences (NINS), 462-2, Nobeyama, Minamimaki, Minamisaku, Nagano 384-1305, Japan}
\altaffiltext{3}{Institute for Advanced Research, Nagoya University, Furo-cho, Chikusa-ku, Nagoya 464-8601, Japan}
\email{ohama@a.phys.nagoya-u.ac.jp}
\email{mikito@a.phys.nagoya-u.ac.jp}
\author{Mikito \textsc{Kohno}\altaffilmark{1}}
\author{Keisuke \textsc{Hasegawa}\altaffilmark{1}}
\author{Kazufumi \textsc{Torii}\altaffilmark{2}}
\author{Atsushi \textsc{Nishimura}\altaffilmark{1}}
\author{Yusuke \textsc{Hattori}\altaffilmark{1}}
\author{Takahiro \textsc{Hayakawa}\altaffilmark{1}}
\author{Tsuyoshi \textsc{Inoue}\altaffilmark{1}}
\author{Hidetoshi \textsc{Sano}\altaffilmark{1,3}}
\author{Hiroaki \textsc{Yamamoto}\altaffilmark{1}}
\author{Kengo \textsc{Tachihara}\altaffilmark{1}}
\author{Yasuo \textsc{Fukui}\altaffilmark{1,3}}%

%%% end:list of authors

%% `\KeyWords{}' always has to be placed before `\maketitle'.
 %Do NOT move this preamble from here!

\maketitle

\begin{abstract}
Understanding the mechanism of O star formation is one of the most important issues in current astrophysics.
 It is also an issue of keen interest how O stars affect their surroundings and trigger secondary star formation. 
 An H\,\emissiontype{II} region RCW 79 is one of the typical Spitzer bubbles alongside of RCW 120.
 New observations of CO $J=$ 1--0 emission with Mopra and NANTEN2 revealed that molecular clouds are associated with RCW 79 in four velocity components over a velocity range of 20 km s$^{-1}$.
 We hypothesize that two of the clouds collided with each other and the collision triggered the formation of 12 O stars inside of the bubble and the formation of 54 low mass young stellar objects along the bubble wall.
 The collision is supported by observational signatures of bridges connecting different velocity components in the colliding clouds.
 The whole collision process happened in a timescale of $\sim$3 Myr.
 RCW 79 has a larger size by a factor of 30 in the projected area than RCW 120 with a single O star, and the large size favored formation of the 12 O stars due to the larger accumulated gas in the collisional shock compression.

\end{abstract}
\section{Introduction}
O-star formation and its influence on triggering secondary star formation are important issues in understanding galaxy evolution because the O stars are highly influential in controlling the physical states of the interstellar medium (ISM) by injecting a large amount of kinetic energy in stellar winds, ultraviolet photons and supernova explosions. H\,\emissiontype{II} regions are an unambiguous signpost of O star formation and deserve an intensive study of their associated molecular gas, the raw material of star formation.	
\par	
RCW 79 is an H\,\emissiontype{II} region listed as an emission nebula by Rodgers et al. (1960) and is named S145 in the Spitzer bubble catalogue (Churchwell et al. 2006). It shows a typical bubble shape similar to RCW 120 (Zavagno et al. 2007; Torii et al. 2015) and is located toward $(l,b)=$(\timeform{308.63D}, \timeform{0.58D}) at a kinematic distance 4.0 kpc in the Crux-Scutum Arm (Russeil et al. 1998). 
Martins et al. (2010) estimated the age of each O-stars from near infrared spectroscopic observations (see Martins et al. 2010 Figure 7 left). They showed that the age of the 12 O-stars is 2.0--2.5 Myr with a typical uncertainty of $\sim$ 0.5 Myr from a Chi-square fitting (see Martis et al 2010 Figure 8 and section 4.2.2). The error bar is too large for estimating the individual age separately. Therefore, we adopted the age of all the O-stars as 2.3$\pm$0.5 Myr.
The radius of the bubble is 7.6 pc (Deharveng et al. 2010). RCW 79 has been a subject of extensive observational studies particularly at infrared wavelengths with Spitzer and Herschel (Zavagno et al. 2006; Liu et al. 2017). A picture which was discussed in the previous papers is that the central O stars compress the ambient molecular gas into a shell and low-mass stars were formed as the second generation stars in the shell. Liu et al. (2017) presented an extensive work on the identification of dust clumps mainly based on the recent Herschel data and identified 54 clumps in the bubble wall. They include Class0, Class0/I and Class I objects, which is estimated to be 0.5 Myr (Liu et al. 2017). These authors claimed that the sample represents young stars formed by ionization-driven compression around the H\,\emissiontype{II} region. As such, it was thought that RCW 79 is an example of triggered second-generation star formation in an expanding bubble driven by an H\,\emissiontype{II} region (Zavagno et al. 2006). A CO $J=$ 3--2 survey for 43 Spitzer bubbles were made by Beaumont and Williams (2010) and these authors showed that the CO clouds associated with the bubbles have no sign of spherical expansion, but has a general morphology which is plane-like with a hole. The results do not seem to immediately support the collect-collapse scenario.

Another typical example of a bubble is RCW 120 ionized by a central O 7.5 star (for review see Deharveng and Zavagno 2008). It is argued that triggered formation of second-generation low-mass stars is taking place in the RCW 120 shell based on infrared and CO observations (Zavagno et al. 2007), while the formation of the central O star was not pursued in their scenario. Recently, RCW 120 was extensively studied in the CO emission lines by Torii et al. (2015). These authors found that the main molecular cloud shows a good correspondence with the ring-like distribution of the infrared nebula, whereas they did not see a sign of expansion of the molecular gas. The expansion is a natural outcome if it was formed by gas acceleration due to the central O star. These authors found that another cloud with a velocity separation of 20 km s$^{-1}$ is associated with opening of the bubble as evidenced by the high temperature heated by the H\,\emissiontype{II} region. Since the large velocity cannot be bound by gravity, Torii et al. (2015) lead to an alternative scenario that the two clouds collided with each other $\sim$Myr ago and the small cloud created a cavity in the large main cloud. According to the scenario, an O star was formed by the collisional compression as numerically simulated by Habe and Ohta (1992), and the inside of the cavity is ionized by the O star. These observations illustrate that molecular observations which can discriminate velocity separation is crucial in understanding star formation in these bubbles.

RCW 79 is part of the large-scale dense molecular gas of 100-pc extent as revealed by C$^{18}$O $J=$ 1--0 observations of the Centaurus region with NANTEN (Saito et al. 2001). These observations identified a C$^{18}$O cloud at $-46.8$ km s$^{-1}$ associated with RCW 79 (no.3 in their list) at 2.7 arcmin resolution, which is a small part of the most massive C$^{18}$O aggregation in NANTEN Galactic plane survey (Mizuno and Fukui 2004). The C$^{18}$O cloud mass is estimated to be $1.2\times10^4\ M_{\odot}$, and was shown to be an active site of star formation evidenced by luminous infrared sources. The aim of the present study is to reveal the detailed distribution and kinematics of the molecular gas in RCW 79 at angular resolutions better than arcmin and to explore how star formation is proceeding. This paper is organized as follows; Section 2 describes observational details and Section 3 observational results. Section 4 discusses star formation processes in RCW 79 based on a  cloud-cloud collision scenario and Section 5 concludes the paper.

\section{Observations}
\subsection{NANTEN2 $^{12}$CO ($J=$ 1--0) observations}
We made $^{12}$CO ($J=$ 1--0) observations with the NANTEN2 4 m millimeter/sub-millimeter radio telescope of Nagoya University. The observations of $^{12}$CO ($J =$ 1--0) emission were carrieded out from May 2012 to December 2012. The front end was a 4 K cooled Superconductor-Insulator-Superconductor (SIS) mixer receiver. The system temperature including the atmosphere was $\sim$ 120 K in the double-side band (DSB) including the atmosphere toward the zenith. 
The backend was a digital-Fourier transform spectrometer (DFS) with 16384 channels of 1 GHz bandwidth. The velocity coverage and resolution were $\sim 2600$ km s$^{-1}$ and 0.16 km s$^{-1}$, respectively. We used the on-the-fly (OTF) mapping mode. The pointing accuracy was confirmed to be better than \timeform{20"} with daily observations toward the Sun. 
The absolute intensity calibration was applied by observing IRAS 16293−2422 [$\alpha_{\rm J2000} = \timeform{16h32m23.3s} , \delta_{\rm J2000} = \timeform{-24d28'39. 2"}$] (Ridge et al. 2006). The final beam size was \timeform{180"} (FWHM). The typical rms noise level was $\sim$ 0.59 K ch$^{-1}$ with a velocity resolution of 0.95 km s$^{-1}$.

\subsection{Mopra $^{12}$CO, $^{13}$CO and C$^{18}$O ($J=$ 1--0) observations}
We used the 22 m ATNF (Australia Telescope National Facility) Mopra telescope in Australia from June to August in 2013. 
The half power beam width (HPBW) was $\sim$ \timeform{30"} at 115 GHz. 
The typical system noise temperature ($T_{\rm sys}$) was 400--700K in the single-side band (SSB).
The backend system “MOPS” has 4096 channels across 137.5 MHz in each of the two polarizations, and the velocity resolution was 0.088 km s$^{-1}$ and the velocity coverage was 360 km s$^{-1}$ at 115 GHz.
We used the OTF mapping mode, and {an} observed area was \timeform{0.3D} $\times$ \timeform{0.6D}.
The {pointing} accuracy was checked to be within \timeform{6"} by observing the SiO maser sources (W Hya and R Car).
The intensity calibration was applied by observing {M17SW [$\alpha_{\rm J2000} =\timeform{18h17m30.0s}, \delta_{\rm J2000} = \timeform{-16D13'06.0"}$]. 
The rms noise level was $\sim$ 0.75 K ch$^{-1}$ for $^{12}$CO $(J =$ 1--0), $\sim$ 0.53 K ch$^{-1}$ for $^{13}$CO $(J =$ 1--0), and $\sim$ 0.64 K ch$^{-1}$ for C$^{18}$O $(J =$ 1--0) with a velocity resolution of 0.26 km s$^{-1}$.
The data cube was smoothed with a Gaussian kernel of \timeform{30"}, and the final beam size was $\sim$ \timeform{40"} (FWHM).
%\subsection{ASTE $^{12}$CO ($J=$ 3--2) observations}
% We carried out $^{12}$CO ($J=$ 3--2) observations with the Atacama {Sub-millimeter Telescope Experiment (ASTE, Ezawa et al. 2004, 2008) in June 2014. The frontend was 2SB SIS mixer called “CATS 345 ” (Inoue et al. 2008).} The typical system temperature was {300 K at 345 GHz} in the single-side band (SSB). The back end was the DFS, “MAC” (Sorai et al. 2000), with 1024 channels of 128 MHz bandwidth. The {pointing} accuracy was checked to be within \timeform{3"}. The intensity calibration was applied by observing {IRC10216 [$\alpha_{\rm J2000} =\timeform{09h45m14.8s}, \delta_{\rm J2000} = \timeform{13D30'40.0"}$]  and W28 [$\alpha_{\rm J2000} =\timeform{17h57m26.8s}, \delta_{\rm J2000} = \timeform{-24D03'54.0"}$]. The data cube has a beam size of $\sim$ \timeform{28"} and the rms noise level was $\sim$ 0.10 K ch$^{-1}$ for $^{12}$CO $(J =$ 3--2) with a velocity resolution of 0.53 km s$^{-1}$.   

\subsection{Archive datasets}
We used the following archive datasets to compare with the CO data. i.e., the H$\alpha$ optical data from SuperCOSMOS (Parker et al. 2005) observed with the UK Schmidt Telescope (UKST) of the Anglo-Australian Observatory; near and mid-infrared data from Spitzer space telescope (GLIMPSE in 8.0 $\mu$m, Benjamin et al. 2003, MIPSGAL in 24 $\mu$m, Carey et al. 2009); and the 843 MHz radio continuum emission data from SUMSS (the Sydney University Molonglo Sky Survey : Bock et al. 1999, Mauch et al. 2003) observed with the Molonglo Observatory Synthesis Telescope (MOST).

\section{Results}
\subsection{Overall CO distribution}
Figure 1 shows a composite infrared image of RCW 79 taken with Spitzer ({GLIMPSE : Benjamin et al. 2003, MIPSGAL : Carey et al. 2009}), where another spitzer bubble S144 is indicated. 

Figure 2 shows $^{12}$CO $J =$ 1--0 images obtained with NANTEN2 toward RCW 79 in velocity- channel distributions covering 34 km s$^{-1}$ every 2.9 km s$^{-1}$.
 The CO emission ranges from $-64.4$ km s$^{-1}$ to $-30.1$ km s$^{-1}$.
 We find that the emission at $-50.1$ km s$^{-1}$ $\textless$ $V_{\rm LSR}$ $\textless$  $-41.5$ km $^{-1}$ has three peaks along the bubble, and shows best correspondence with the bubble among the various velocity components. 

Figures 3a, 3b and 3c show three components at [$-60.1$ km s$^{-1}$ $\textless$ $V_{\rm LSR}$ $\textless$ $-55.1$ km s$^{-1}$], [$-55.1$ km s$^{-1}$ $\textless$ $V_{\rm LSR}$ $\textless$ $-45.1$ km s$^{-1}$], and [$-45.1$ km s$^{-1}$ $\textless$ $V_{\rm LSR}$ $\textless$ $-37.2$ km s$^{-1}$] in the higher resolution CO images obtained with Mopra.
 The latter two velocity ranges in Figures 3b and 3c show correspondence with the bubble, while the velocity component in Figure 3a show no apparent correspondence.
 The most prominent features are seen at $-47$ km s$^{-1}$ toward the bubble and have three peaks which are denominated P, Q and R for convenience (Figure 3b).
 In Figures 3a and 3b the CO emission is found in the north of the bubble, where the bubble shows an opening, and a strip in an X-Y coordinate is defined by the two dashed lines.
 Velocity-channel distributions of $^{12}$CO with Mopra are shown at a finer 1 km s$^{-1}$ velocity interval in the Appendix.
 Figure 3d shows a position-velocity diagram along the Galactic latitude. The most intense component is at $-47$ km s$^{-1}$, and the secondary component is peaked at $-40$ km s$^{-1}$.
 We find connecting features between the $-47$ km s$^{-1}$ component and the $-40$ km s$^{-1}$ component at $b$ less than \timeform{0.55D}, and at $b= \timeform{0.7D}$ -- \timeform{0.75D}.
 Connection between the $-57$ km s$^{-1}$ and the $-47$ km s$^{-1}$ components is not clear in the figure.
 The velocity distribution in the main peak is flat with no significant gradient or broadening toward the bubble, except for the peak at $b=\timeform{0.53D}$, which shows a broad linewidth of 4 km s$^{-1}$.
 The rest of the emission of the main component has a narrower linewidth of 2--3 km s$^{-1}$.
Figure 4 shows detailed velocity distribution in the northern part of the bubble in Figures 4a, 4b, 4c, and 4d.
 This part mainly includes the $-57$ km s$^{-1}$ component located in the north of the bubble.
 We define X- and Y-axes normal to and along the dashed lines.
 Figure 4e shows a position-velocity diagram along the Y-axis which includes an entire velocity range of the emission in Figure 4.
 The strongest emission is seen at $-47$ km s$^{-1}$, and the secondary component at $-57$ km s$^{-1}$.
 Between the two components, we find another component at $-52$ km s$^{-1}$ which shows highly filamentary distribution; in Figure 4c we find that five filamentary features which are parallel with each other and with the X axis as indicated by arrows. 
 The five are seen at $-53$ km s$^{-1}$ in Figure 4e. We note that there are bridge features connecting the $-47$ km s$^{-1}$ and $-52$ km s$^{-1}$ components at three directions and that connecting the $-52$ km s$^{-1}$ and $-57$ km s$^{-1}$ components at two directions. 
 These connecting features are suggestive of physical linkage among all these features in Figure 4 as discussed later in Section 4.
 
 The distance of the cloud was estimated to be 4.5 kpc for cloud velocity of $-47$ km s$^{-1}$ by using the Galactic rotation model (Saito et al. 2001).
 Russeil et al. (1998) derived 4.0 kpc by H$\alpha$ velocity of $-49$ km s$^{-1}$.
 In the present paper, we adopt 4.5 kpc which corresponds to the kinematic distance of the main cloud associated with RCW 79.
 The total cloud mass was calculated to be $6 \times 10^4 M_{\odot}$ by adopting an XCO factor of $1.0\times10^{20}$ cm$^{-2}$ /K km s$^{-1}$ in Perseus (Okamoto et al. 2017). 
 Masses of the individual components are estimated as follows; masses of P, Q, R, and the -57 km s$^{-1}$ cloud are calculated to be $2 \times 10^4 M_{\odot}$, $7 \times 10^3 M_{\odot}$, $1 \times 10^4 M_{\odot}$, and $6 \times 10^3 M_{\odot}$, respectively.

\subsection{Comparison with other wavelengths}
Figure 5 shows comparison between images of two clouds and other four wavelengths. 
 Figure 5a and 5e show that the 8 $\mu$m emission, delineates the inner surface of the bubble. 
 The 8 $\mu$m band contain thermal emission from hot dust plus emission from Polycyclic Aromatic Hydrocarbon (PAH) features 7.7 and 8.6 $\mu$m (Drain 2003, Churchwell et al. 2004). 
 Figures 5b and 5f show the 24 $\mu$m emission of the heated dust in the bubble, and the ionized gas observed in thermal radio emission and H$\alpha$ emission (Figure 5c and 5g). 
 The radio emission which is not affected by extinction shows that the strongest emission is seen close to the O stars due to intensive ionization of the molecular gas (Figure 5d and 5h).
 The radio continuum emission extends beyond the bubble to the south, indicating that part of the ionized gas is escaping from the bubble.
 The  $-57$ km s$^{-1}$ cloud is distributed outside of bubble (Figure 5 (a-d)). The CO intensity peaks of the $-47$ km s$^{-1}$ cloud coincide with 8 $\mu$m infrared emissions (Figure 5e), on the other hand, the 24 $\mu$m, 843 MHz, and H$\alpha$ emissions are bright inside the two peaks of clouds (P, Q) in RCW 79 (Figure 5f, g, and h).
  Figure 6 shows 54 YSO candidates (Liu et al. 2017) superposed on the $-57$ km s$^{1-}$ and $-47$ km s$^{1-}$ cloud. 
 These YSOs show good correspondence with the CO peaks P, Q, and R, indicating that they were formed in the clouds (Figure 6b).
 The first moment map indicates that the $-57$ km s$^{-1}$ and  $-47$ km s$^{-1}$ cloud are separated, and that the $-57$ km s$^{-1}$  cloud is located in the north and outside of the bubble (Figure 6c).
 %Figure 6 shows 54 YSO candidates (Liu et al. 2017) superposed on the CO main component.
 %These YSOs show good correspondence with the CO peaks P, Q, and R, indicating that they were formed in the clouds.

\section{Discussion}
\subsection{Formation mechanisms of the bubble}
RCW 79 attracted attention as a candidate site of triggered star formation under the effect of H\,\emissiontype{II} regions (Deharveng et al. 2010; Zavagno et al. 2006 etc.).
 A recent paper based on accumulated infrared data with Herschel, archival 2MASS, Spitzer and WISE data provide significant details of the candidates for young low-mass stars in RCW 79 observed as dust clumps, whereas large obscuration hampers to derive their stellar properties into detail (Liu et al. 2017).
 These clumps show strong spatial correlation with the bubble-like distribution of gas and dust as shown in Figure 6; for each clump, size is estimated to be 0.1--0.4 pc, density (0.1--44)$\times 10^5$ cm$^{-3}$, and infrared luminosity (19--12712) $L_{\odot}$ (Liu et al. 2017).
 There is also an O star in a small Spitzer bubble S144 toward the peak P. 
 These clumps seemingly offer an ideal site to study the second-generation star formation in the collect-collapse scenario (Zavagno et al. 2006). 

In the literature, it is usually argued that these stars are formed under triggering by the H\,\emissiontype{II} region following the collect-collapse scenario.
 If efficient acceleration of the ambient molecular gas by the H\,\emissiontype{II} region is confirmed, this is a possible scenario to explain the bubble shape and star formation therein.
 It is however to be noted that detailed theoretical modeling is missing to confront observations with theories in the triggered formation/collapse of the clumps.
 It is also an issue which is to be understood how the first generation O star(s) were formed in isolation as an initial step of the bubble formation; in RCW 79, the 12 O stars are located inside the bubble.
 This was not explored in the published works on the second-generation star formation, and a picture of star formation in RCW 79 remains incomplete.

 In order to better understand the star formation mechanism in bubbles we shall test observationally other possible mechanisms of bubble formation.
 In RCW 120, Torii et al. (2015) did not find evidence for expansion of the bubble shell in CO and argued that the collect-collapse scenario which predicts shell expansion by the central O star(s) is not supported.
 These authors presented an alternative scenario on formation of RCW 120 based on a cloud-cloud collision.
 In this scenario, a small cloud and a large cloud collided at a velocity of $\sim$ 20 km s$^{-1}$ and the small cloud created a cavity in the large cloud.
 The interface layer becomes strongly compressed by the supersonic collision to form an O star which is now ionizing the H\,\emissiontype{II} region inside the cavity.
 This scenario is supported by theoretical studies.
 The cloud morphology where a cavity is formed by collision was numerically simulated by Habe and Ohta (1992), Anathpindika (2010), and Takahira et al. (2014), and it was demonstrated that a similar cavity as observed is reproducible by a cloud-cloud collision.
 The physical conditions of the interface layer is calculated by magneto-hydrodynamical (MHD) numerical simulations by Inoue and Fukui (2013), and it is shown that the interface layer becomes dense and highly turbulent; in the layer the collision increases the sound speed by a factor of 5 and the mass accretion rate by two orders of magnitude which is proportional to the third power of the sound speed.
 The mass accretion rate becomes as high as $2 \times 10^{-4}\ M_{\odot}$ yr$^{-1}$ (Inoue \& Fukui 2013; Inoue et al. 2018), which satisfies the threshold value required for formation of a $\sim$ 100 $M_{\odot}$ star (Wolfire and Cassinelli 1987). 

\subsection{A cloud-cloud collision scenario in RCW 79}
RCW 79 is associated with the shell-like main molecular cloud consisting of three peaks along the infrared bubble (Figure 3b). The main cloud shows no sign of expansion (Figure 3d).
 Another molecular cloud blue-shifted by 12 km s$^{-1}$ is located just outside the opening of the bubble, and shows a sign of dynamical interaction as the bridge features with parallel streamers.
 The secondary cloud is physically interacting with the main cloud as proved by the bridge features accompanying an intermediate velocity layer, whereas the cloud is distributed outside the bubble.
 Figure 4f presents a snap shot of two colliding molecular flows by MHD numerical simulations (Inoue and Fukui 2013). 
 This image shows a position-velocity diagram taken in the moving direction of the head-on colliding molecular flows, and illustrates the behavior of the interface layer and the bridge features.
 The main parameters of the numerical model are given in Model 2 of Table 1 in Inoue and Fukui (2013).
 Torii et al. (2015) argued that the remnant of the secondary cloud outside the opening of the bubble is consistent with the numerical simulations of two colliding clouds of Habe and Ohta (1992) if we consider a realistic cloud shape where the colliding “small” cloud is a tip of a more extended cloud. 
 The bubble-like main cloud and the secondary cloud outside the opening of the bubble were found in RCW 79, and these observed properties of the molecular gas are common to RCW 120 where a cloud-cloud collision scenario was presented.
 Based on the present results and reasoning, we set up a cloud-cloud collision scenario in RCW 79.
 The outline of the scenario is as follows; “a small cloud ”collided with “a large cloud” and produced a cavity in the large cloud; the size of the small cloud, $\sim$ 15 pc in diameter,  was slightly smaller than the collided part of the large cloud and the interaction left three clouds along the bubble as observed P, Q, and R which reflected the outer shape of the large cloud.
 We are not able to see the small cloud inside the bubble at present, because the small cloud is already dissipated by the collisional interaction and ionization.
 
% \textcolor{red}{RCW 79では母体分子雲がすでに散逸されているため、分子雲衝突を観測的に直接に示すことは困難である。しかし、我々のこれまでの研究から10個を超えるO型星を含むスーパースタークラスターのうち、母体分子雲を含む天体は、分子雲衝突が関係していることを突き止めた (Furukawa et al. 2009, Ohama et al 2010, Fukui et al. 2014, 2016)。また、天の川銀河にある20個以上の大質量星形成領域について、分子雲衝突を示した (ref)。これらの統計的な研究結果から12個のO型星を含むRCW 79 についても分子雲衝突による形成シナリオを提案することに大きな矛盾はないと考える。また、RCW 79ではすでにO型星が形成されているので、電離の影響も存在するが、衝突と電離の影響を現在の観測結果のみから切り分けることは困難である。}
 
 In the collisional interface layer the 12 O stars were formed inside the bubble.
 The 12 O stars are not accompanied by lower mass stars which are common in known stellar clusters. 
 The dense gas is already ionized and we do not see directly the formation of the O stars. By MHD simulations Inoue and Fukui (2013) showed that the collision compressed layer forms dense and massive cores which are likely precursor of O stars. 12 O stars may require molecular cloud core mass exceeding $\sim$ 400 $M_{\odot}$ ($12\times30$ $M_{\odot}$). The massive core is likely produced in the present large-scale collision which includes large cloud mass of $10^4$ $M_{\odot}$. It is probable that the O star formation is correlated with the molecular mass. The O stars are isolated from the low mass members in RCW 79, implying a top-heavy core mass function, which is characteristic to collisional compression (Inoue and Fukui 2018, in preparation). Ionization will disperse the cloud core just after the O star formation, whereas the rapid mass accretion of $2 \times 10^{-4}$ $M_{\odot}$ yr$^{-1}$ probably allowed the 12 O star formation within a few times $10^4$ yrs. In other cases of cloud-cloud collision, RCW 120 and M20, only a single O star was formed. This likely reflects the small mass of these clouds, $\sim 10^3$ $M_{\odot}$ (Torii et al. 2011,and 2016).}
 %This indicates an extreme top-heavy stellar mass function inside RCW 79. We suggest that a top-heavy core mass function was created in the collisionally compressed molecular gas inside the bubble, leading to the 12 O stars with no low mass stars. The numerical simulations of collision by Inoue and Fukui (2013) indicate formation of heavy core mass in the shocked layer and lend support for the top-heavily core mass function.
 The O-star formation took place near the symmetry axis of the head-on collision when the accumulated gas became large enough for O star formation, and this happened inside the current bubble.
 A schematic of the collision of the two clouds is shown in Figure 7. 
%In the present scenario, “the small cloud" here refers to only the collisional interacting tip of a more extended cloud. In the large scale view of CO (Figure 2), we suggest that “the small cloud” corresponds to a tip of the 57 km s$^{-1}$ cloud as indicated by an arrow in Figure 2; the 57 km s$^{-1}$ cloud is extended normal to the Galactic plane by $\sim$100 pc $\times \sim$30 pc (see also for RCW 120 Figure 2 of Torii et al. 2015). 
 The mass of the -47 km s$^{-1}$ cloud is estimated to be $2\times 10^4 M_{\odot}$, and that of the -57 km s$^{-1}$ cloud $5\times 10^3 M_{\odot}$ at a contour level of 8 K km s$^{-1}$ and 14 K km s$^{-1}$, respectively in Figure 2.
In Figure 6d, the dashed line shows the projected outer boundary of the cylindrical cavity created by the collision in the $-47$ km s$^{-1}$ cloud.
% In Figure 6 we show a possible projected path of the center of the small cloud by an arrow, where we assume that the path is straight.
 The O star formation happened close to the current position, while the location of the O stars shows a shift from the line by $\sim$ 2 pc to the west.
 It is possible that the O stars moved along the line by several pc to the south at a velocity smaller than the original collision velocity by deceleration due to momentum conservation (Haworth et al. 2015ab).
 The number of O stars in RCW 79 is ten times larger than that of RCW 120 and other Spitzer bubbles which have usually a single O star.
 We suggest that the large number of O stars is due to a larger size of the cavity which allowed the interface layer to accumulate larger molecular mass than in RCW 120.
 If a mass accretion rate of $2 \times10^{-4}\ M_{\odot}$ yr$^{-1}$ is assumed, the time scale of O star formation is estimated to be 20 $M_{\odot}/2 \times 10^{-4}\ M_{\odot}\ {\rm yr}^{-1} \sim 0.1$ Myr. 
 
% \textcolor{red}{我々は、RCW 79 を構成する星の年齢を考慮に入れた以下のような分子雲衝突モデルを提案する。
% \begin{itemize}
%\item 3 Myr ago : -47 km/s cloudと-57 km/s cloudが相対速度 10 km/s で衝突。
%\item 2 Myr ago : 境界面で圧縮層が形成され10 pc 進んだところで、12個のO型星が形成。衝突のタイムスケールは、10 pc/10 km/s 〜 1 Myr となる。
%\item 0.5 Myr ago : さらに8 pc 進んだところで、衝突が終了。バブルの縁でClass 0/I YSOの形成が開始。衝突のタイムスケールは、衝突雲が減速を受けて相対速度が半分になったと仮定すると、8 pc/5 km/s 〜1.6 Myr　となる。
 %\end{itemize}
%この衝突モデルはRCW 120(Torii et al. 2015)の場合と比較すると、バブルのサイズが大きく、衝突のタイムスケールも長い。RCW120よりも衝突による圧縮が長時間起きたことによって、O型星が12個形成されたと考えられる。
%この衝突モデルを導入することで、これまでの先行研究で示唆されている星の年齢の違いも含めて矛盾なく説明することができる。}

We present a scenario of the cloud-cloud collision by considering the stellar ages estimated in the literature, where we adopt length and velocity for an assumed tilt angle of 45 degrees of the relative motion to the line of sight. The angle, while not estimated observationally, takes into account the large observed velocity difference $\sim$10 km s$^{-1}$ which suggests that the tilt angle is not close to 0 degrees. The following discussion is not much dependent on the tilt angle. The age of the 12 O stars is estimated to be $2.3\pm0.5$ Myr (Martins et al. 2010). Liu et al. (2017) observed 22 sources in the infrared condensations of RCW 79, and considered that they are low mass young stars, Class 0/I YSOs. These authors adopted their age to be $\sim 0.5$ Myrs (Liu et al. 2017; also Evans et al. 2009 and Enoch et al. 2010). From the observed velocity and size of the bubble, we are able to estimate the time evolution of the collisional process. The colliding clouds are extended by $\sim$20 pc (Figure 6), and the collision consists of three different stages as follows. Figure 7 depicts the three stages as explained below;
\begin{itemize}
\item Stage I: The small cloud started collision with the large cloud $\sim$ 2.6 Myrs ago at a relative velocity of $\sim$ 15 km s$^{-1}$, and continued to create a cavity in the large cloud. The interface layer between the two clouds formed and grew in mass.
\item Stage II: The interface layer became dense by compression and O stars form by the gravitational instability in the layer $\sim$ 2.3 Myrs ago. The O star formation happened at a mass accretion rate of $2 \times 10^{-4}\ M_{\odot}$ yr$^{-1}$ within $10^5$ yrs, where the accretion rate is adopted from the numerical simulations by Inoue and Fukui (2013). The O stars ionized their surroundings in the northwestern half of the interface layer and the mass growth of the O stars was terminated by the ionization. The O stars moved to the present position in 2.3 Myrs after travelling $\sim 15$ pc. The average velocity of the O stars, which held the velocity of the layer, is estimated to be $\sim 6$ km s$^{-1}$.
\item Stage III: The remaining interface layer continued to deepen the cavity and reached $\sim$ 30 pc into the large cloud where the layer is halted by the large cloud 0.5 Myrs ago. The average velocity of the layer is given by a ratio $\sim$ 30 pc/2.6 Myrs $\sim$ 10 km s$^{-1}$. The gas between the layer and the large cloud became compressed at a smaller velocity than in the O star formation and formed low mass stars as well as the exciting star of S144 0.5 Myrs ago in the southern end of the bubble. 
\end{itemize}
Over the three stages, we expect the relative velocity between the two clouds was continuously slowed down due to the momentum conservation (e.g., Haworth et al. 2015a, 2015b). We suggest that the O stars were formed in the dense part of the interface layer that was more heavily decelerated than the rest of the layer, which explains that the O stars are still inside the bubble.

In the present scenario, the natal molecular gas is already ionized in an age of RCW 79 $\sim$2--3 Myrs within several pc of the 12 O stars. It is impossible to trace the collision signatures directly toward the O stars.  In a much younger object like M20 with an age less than 0.5 Myrs (Torii et al. 2017), two clouds colliding with the bridge feature are directly observed thanks to much less ionization. A case similar to RCW 79 is found in another 2 Myr old HII region RCW 32 as reported by Enokiya et al. 2018 (arXiv:171100722), where the central region is highly ionized and complementary colliding clouds are seen only outside the H\,\emissiontype{II} region. Similar colliding clouds outside the H\,\emissiontype{II} region are observed in RCW 79 as shown in Figure 4 of the present paper. The central cluster in RCW 79 consists of only 12 O stars with no significant low mass members and is compact with a size of 1 pc. These cluster properties are unusual as compared with the known open clusters, which have the universal stellar mass function and are more extended by a few pc. The RCW 79 case suggests an external impulsive trigger as its formation mechanism. Considering the strong signatures of two-cloud collision in the north (Figure 4), we suggested that a cloud-cloud collision triggered the formation of the 12 O stars. This scenario is consistent with a general picture of two colliding clouds of different sizes by theoretical numerical simulations (Habe and Ohta 1992; Anathpindika 2010; Takahira et al. 2014). Further, the MHD numerical simulations showed that massive dense cores of a few 100 $M_{\odot}$ are formed in the interface layer as a result of collision, which explains the O-star formation in RCW 79. The proposed scenario of cloud-cloud collision is therefore supported by the observed collision signature of the cloud system and the relevant theoretical works

The other YSOs in Q and R may have been formed in the collision in the peripheral part of the small cloud, where collision was probably oblique. Their formation epoch may be scattered more widely, while it is not constrained better than those in P. An argument which favors the triggering is that we do not see similar YSOs inside the bubble where the pre-existing cloud density was perhaps higher. If the YSOs observed at present were formed prior to the collision, we would find more YSOs inside the bubble.

We note that a similar enhanced star formation on the opposite side of the bubble’s opening is seen in RCW120 (Torii et al. 2015) and in a high mass star formation region AGAL G337.916-00.477 (Torii et al. 2017). It is intriguing to test if a similar trend is found in the other bubbles, which may provide a signature of collision.

 % It is suggested that the large CO clouds in Figure 2 are colliding with each other extensively over a scale of more than 100 pc in Section 3. Nonetheless, it seems that only in several tiny spots high-mass star formation was triggered as seen in the Spitzer bubbles where O/early B stars are ionizing H\,\emissiontype{II} regions. This suggests that only spots of high column density are able to form high-mass stars by collision, if we assume that every Spitzer bubble is a spot of O star formation by a cloud-cloud collision. Figure 2 indicates that $\sim$ 10 O stars are being formed by triggering in a volume of 200 pc $\times$ 100 pc $\times$ 100 pc $= 2\times10^6$ pc$^3$ in a timescale of Myr. An O star formation rate is then estimated to be $\sim 0.1 M_{\odot}$ yr$^{-1}$, which is more than an order of magnitude larger than that estimated from the total star formation rate in the Milky Way, 0.68--1.45 $M_{\odot}$ yr$^{-1}$ for all stellar mass (Robitaille and Whitney 2010) if we assume the total volume $\sim$ $2\times10^{10}$ pc$^3$ of the Galactic disk of 8 kpc radius with 100 pc thickness. This is a significantly large number, suggesting that cloud-cloud collision is an important mode of O star formation, whereas we need to be cautious about a possibility of especially enhanced molecular gas density in the present region which may elevate the collision rate above the Galactic average. This suggests the importance of extending a study of cloud-cloud collision in a significant portion of the whole Galactic disk in future.

\section{Conclusions}
We summarize the conclusions of the present study as follows;
\begin{enumerate}
\item We made new CO observations of the Spitzer bubble RCW 79 with NANTEN2 and Mopra 22m telescope.
 The observations of $^{12}$CO $J=$ 1--0 transition revealed that the bubble is associated with at least four velocity components over a velocity range of 20 km s$^{-1}$.
 One of them, the $-48$ km s$^{-1}$ cloud, shows a good correspondence with the infrared bubble, and the $-58$ km s$^{-1}$ cloud is located at the opening of the bubble in a similar way to RCW120 (Torii et al. 2015).
 We found the other two velocity components at $-53$ km s$^{-1}$ and $-40$ km s$^{-1}$ toward the bubble which are likely associated with the bubble.

\item The above four velocity components are connected by bridge features in velocity, supporting their mutual physical association.
 In particular, the three components in a range from  $-57$ km s$^{-1}$ to $-47$ km s$^{-1}$ show tight mutual connection in the north of the bubble, where the bubble shows the opening.
 This situation is similar to the Spitzer bubble RCW120.
 We interpret that the components in the north just outside of opening of the bubble, represent colliding molecular flows based on the MHD numerical simulations (Inoue and Fukui 2013).

\item We hypothesize that two clouds at --57 km s$^{-1}$ and --47 km s$^{-1}$ collided with each other and the collision triggered the formation of 12 O stars in the bubble and 54 low-mass young stellar objects along the bubble wall in a timescale of 0.5--3 Myr.
 This hypothesis is an alternative to the usual collect-collapse scenario, where the stars are formed as secondary generation stars.
 
 RCW 79 has a larger size than RCW120 by a factor of 30 in the projected area, favoring formation of 12 O stars, which did not happen in a smaller bubble RCW120 where a single O star was formed.
  The 12 O stars are not accompanied by lower mass stars which are common in a usual stellar cluster. This indicates an extreme top-heavy mass function.
 We suggest that the mass function was created by the initial low density of the compressed molecular gas inside the bubble which formed no low mass stars.
The numerical simulation of collision by Inoue and Fukui (2013) lend support for the top-heavily core mass function.

\end{enumerate}

\section*{Acknowledgements}
NANTEN2 is an international collaboration of ten universities: Nagoya University, Osaka Prefecture University, University of Cologne, University of Bonn, Seoul National University, University of Chile, University of New South Wales, Macquarie University, University of Sydney, and Zurich Technical University.
% The ASTE telescope is operated by National Astronomical Observatory of Japan (NAOJ).
The work is financially supported by a Grant-in-Aid for Scientific Research (KAKENHI, No. 15K17607, 15H05694) from MEXT (the Ministry of Education, Culture, Sports, Science and Technology of Japan) and JSPS (Japan Soxiety for the Promotion of Science).

%%%
% See the manual for the detail.
%%%

\clearpage
\begin{figure*}[h]
\begin{center}  \includegraphics[width=17cm]{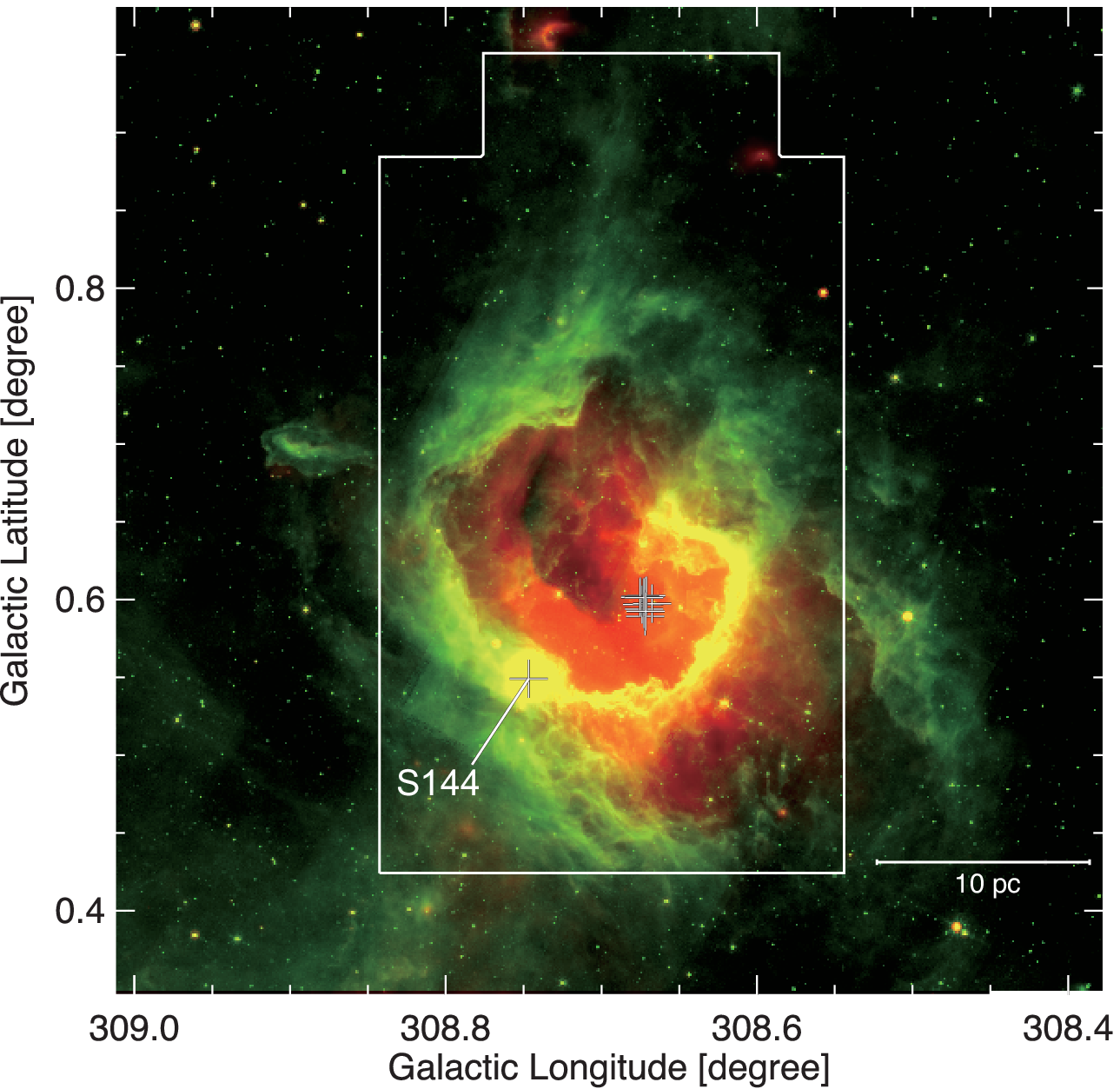}
\end{center}
\caption{Two-color composite images of RCW 79. Green and red show the Spitzer 8 $\mu$m and 24 $\mu$m emission. Crosses depicts exciting O stars (Martins et al. 2010). White highlighted box shows the observed region of $^{12}$CO $J$= 1--0 and $^{13}$CO $J$= 1--0 with Mopra telescope.}\label{.....}
\end{figure*}

\begin{figure*}[h]
\begin{center}  \includegraphics[width=16cm]{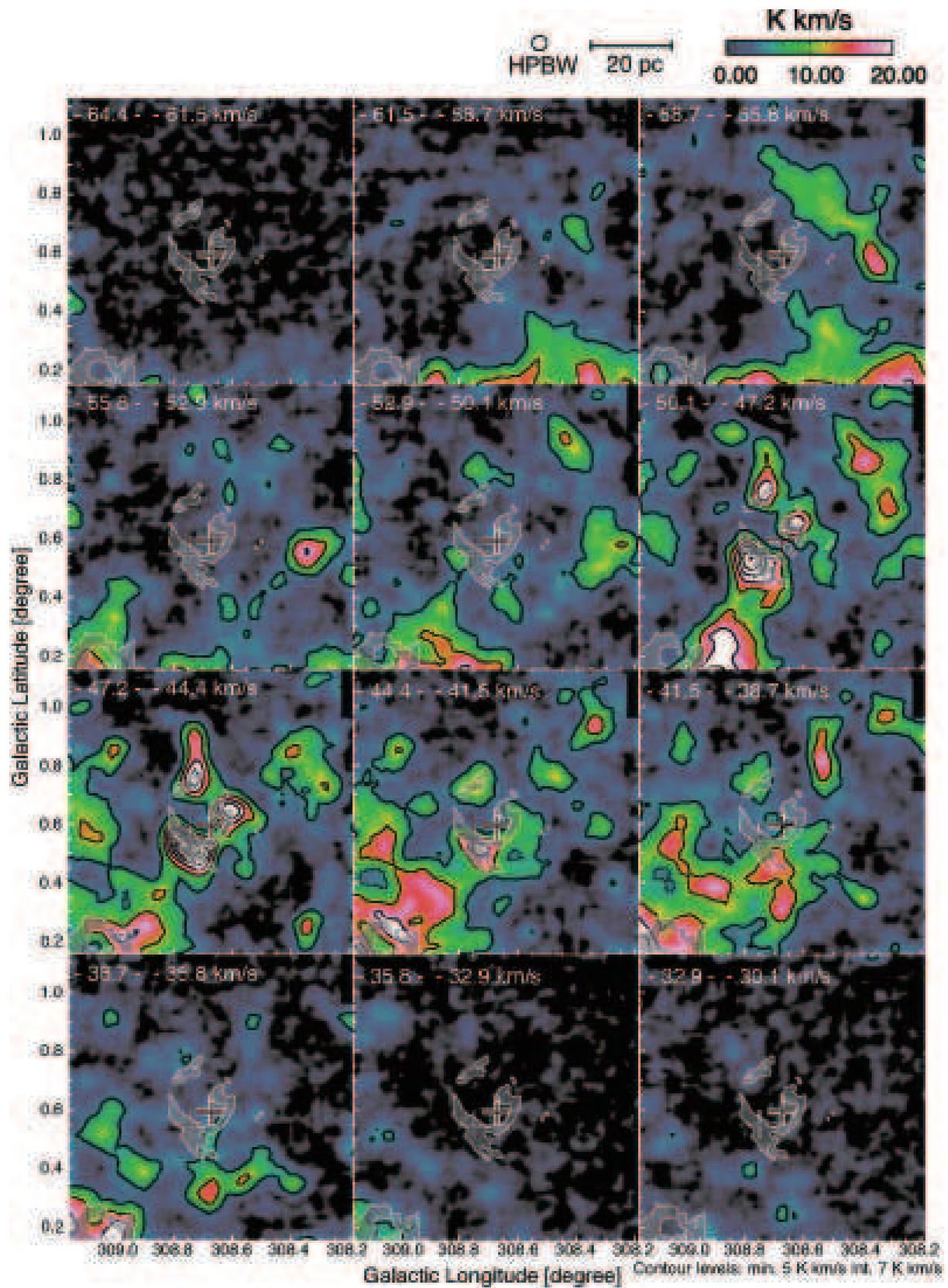}
\end{center}
\caption{NANTEN2 $^{12}$CO $J=$1--0 velocity channel distributions toward RCW 79 at velocity step of 2.9 km s$^{-1}$. Shaded area indicates the 8 $\mu$m ring. Contours are plotted every 6 K km s$^{-1}$ from 6 K km s$^{-1}$. Crosses depicts exciting O stars (Martins et al. 2010).}\label{.....}
\end{figure*}

\begin{figure*}[h]
\begin{center}  
\includegraphics[width=15.5cm]{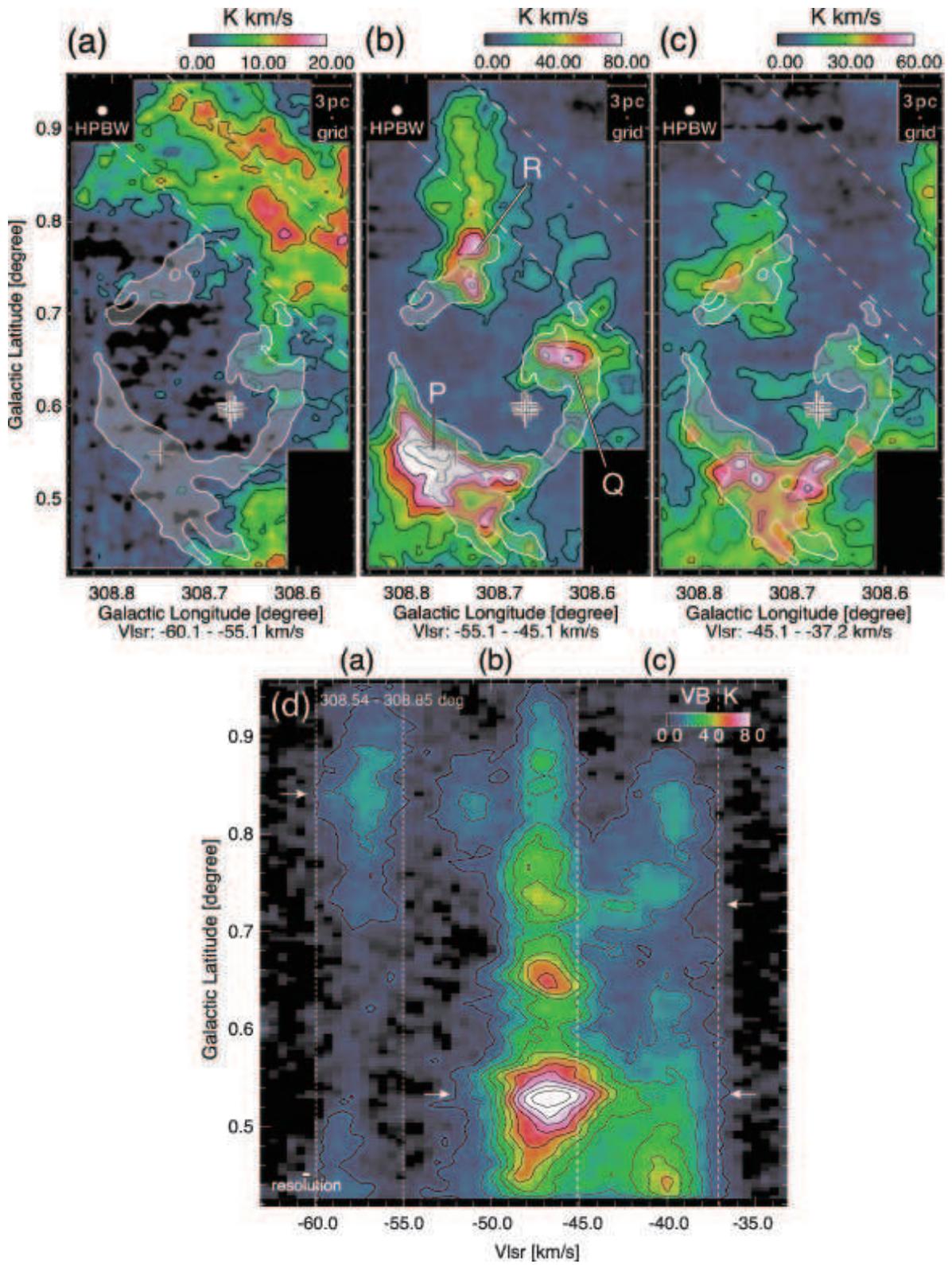}
\end{center}
\caption{$^{12}$CO $J=$ 1--0 distributions toward RCW 79 with the Mopra data. (a) $-60.1$ $\textless$ $V_{\rm LSR}$ $\textless$ $-55.1$ km s$^{-1}$, contours are plotted every 4 K km s$^{-1}$ from 4 K km s$^{-1}$. (b) $-55.1$ $\textless$ $V_{\rm LSR}$ $\textless$ $-45.1$ km s$^{-1}$, contours are plotted 16, 26, 38, 50, 62, 76, 90, 105 K km s$^{-1}$ from 10 K km s$^{-1}$. (c) $-45.3$ $\textless$ $V_{\rm LSR}$ $\textless$ $-35.8$ km s$^{-1}$, contours are plotted every 10 K km s$^{-1}$ from 14 K km s$^{-1}$. Crosses depicts exciting O stars (Martins et al. 2010). Dotted lines show integrated range of Figure 4(e). (d) Latitude-velocity map. Dotted lines and (a-c) indicates integration ranges of (a-c), respectively. Contours are plotted every 0.8 K from 0.8 K}\label{.....}
\end{figure*}

\begin{figure*}[h]
\begin{center}  \includegraphics[width=17cm]{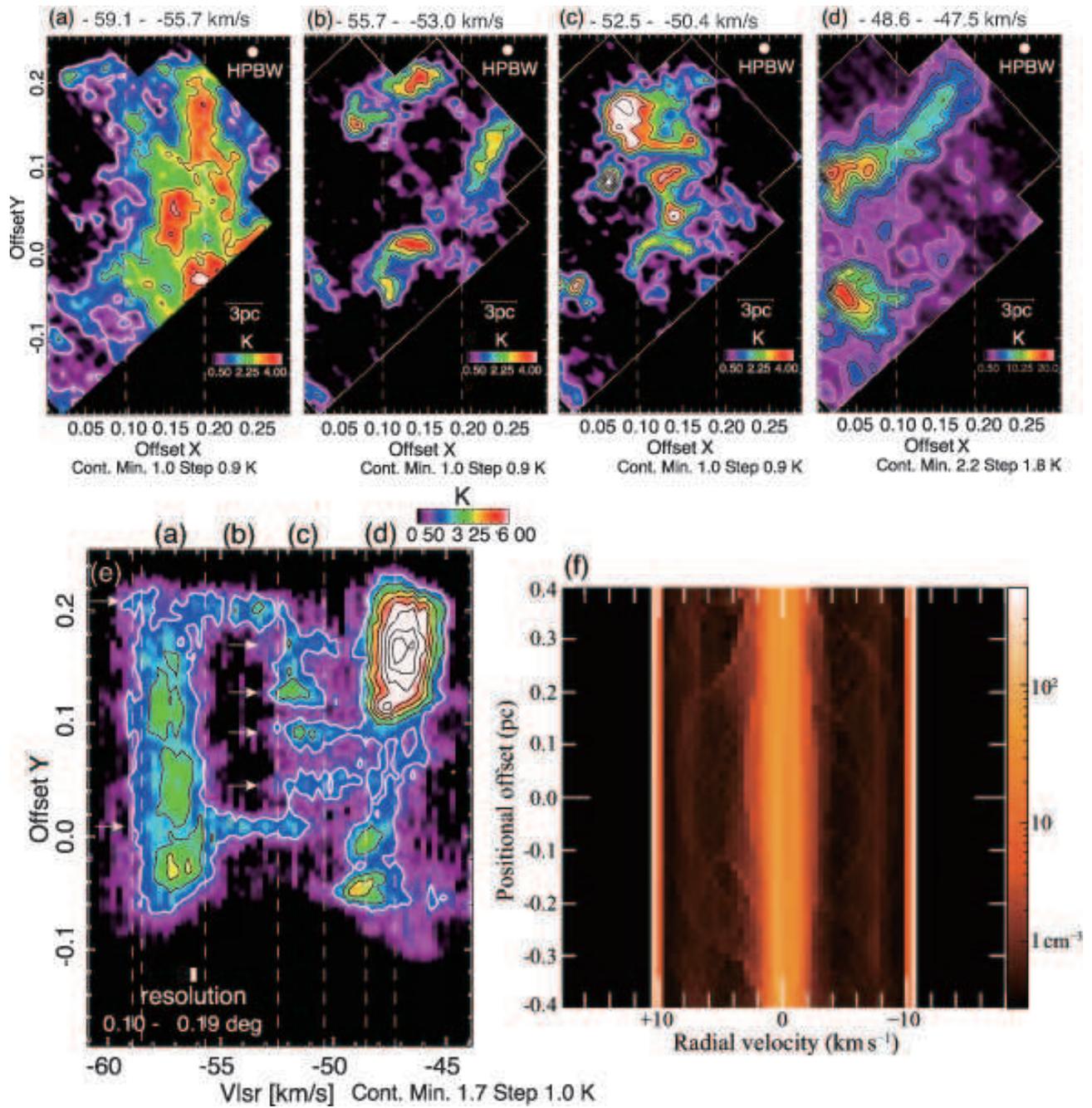}
\end{center}
\caption{(a-d) $^{12}$CO distribution of $-47.5$ degree rotation. Dotted lines are same as that of Figure 3(a-c). (e) Position-velocity map of $-47.5$ degree rotation. The dotted lines show the integrated velocity range of Figure 4(a-d). Dotted lines and (a-d) indicates integration ranges of (a-d), respectively. (f) indicate a snap shot of two colliding molecular flows by MHD numerical simulation (Inoue and Fukui 2013)}.\label{.....}
\end{figure*}

\begin{figure*}[h]
\begin{center}  \includegraphics[width=17cm]{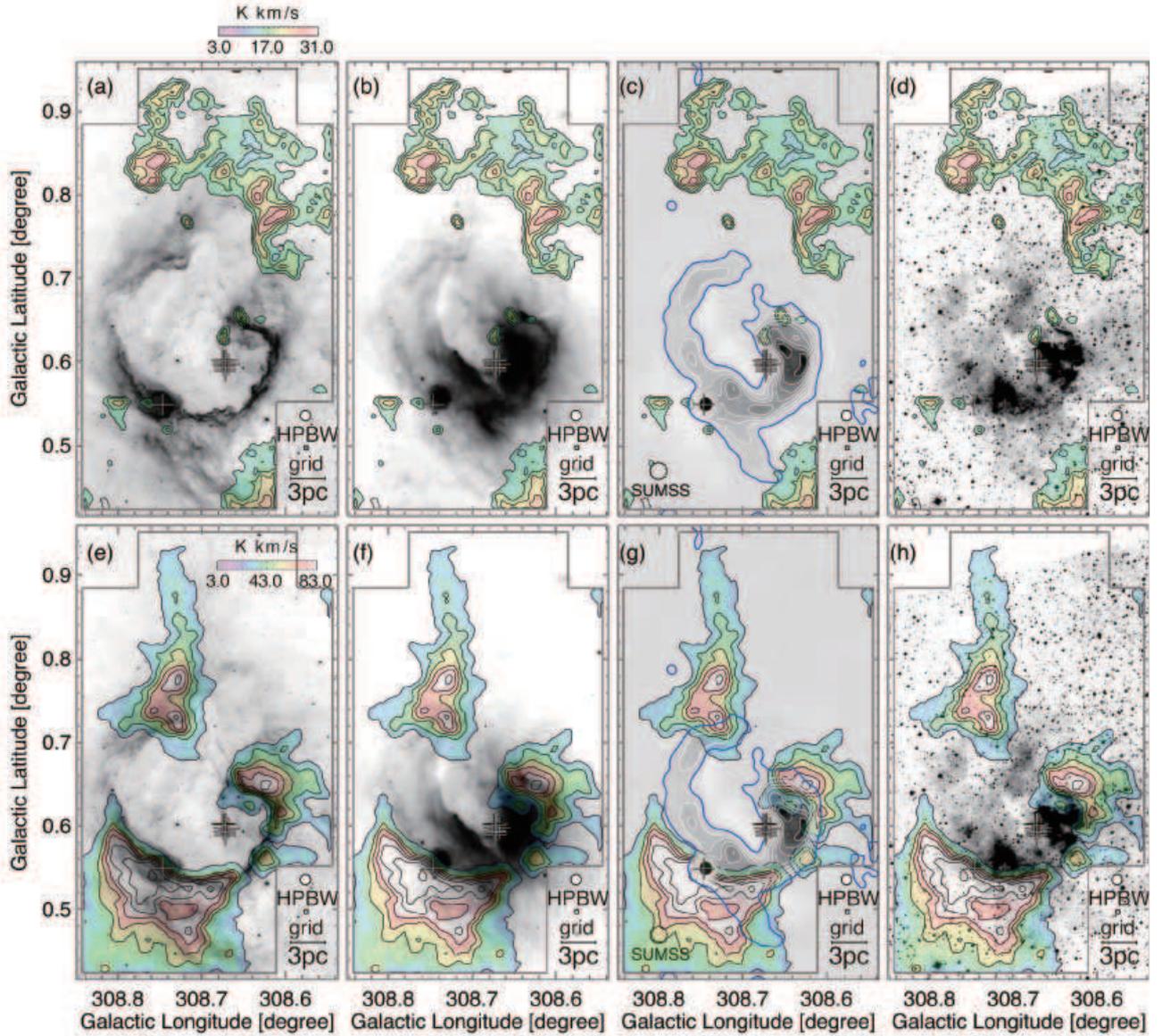}
\end{center}
\caption{(a-d) Comparisons of $-57$ km s$^{-1}$ component (color and contours). (a) the Spitzer 8 $\mu$m emission (grayscale), (b) the Spitzer 24 $\mu$m emission (greyscale), (c) the 843 MHz SUUMSS radio continuum emission (grayscale and thick blue and thin white contours), (d) H$\alpha$ emission (gray scale). (e-h) Comparisons of $-47$ km s$^{-1}$ component (color and contours) in the same manner as (a)-(d).}\label{.....}
\end{figure*}

\begin{figure*}[h]
\begin{center}  \includegraphics[width=17cm]{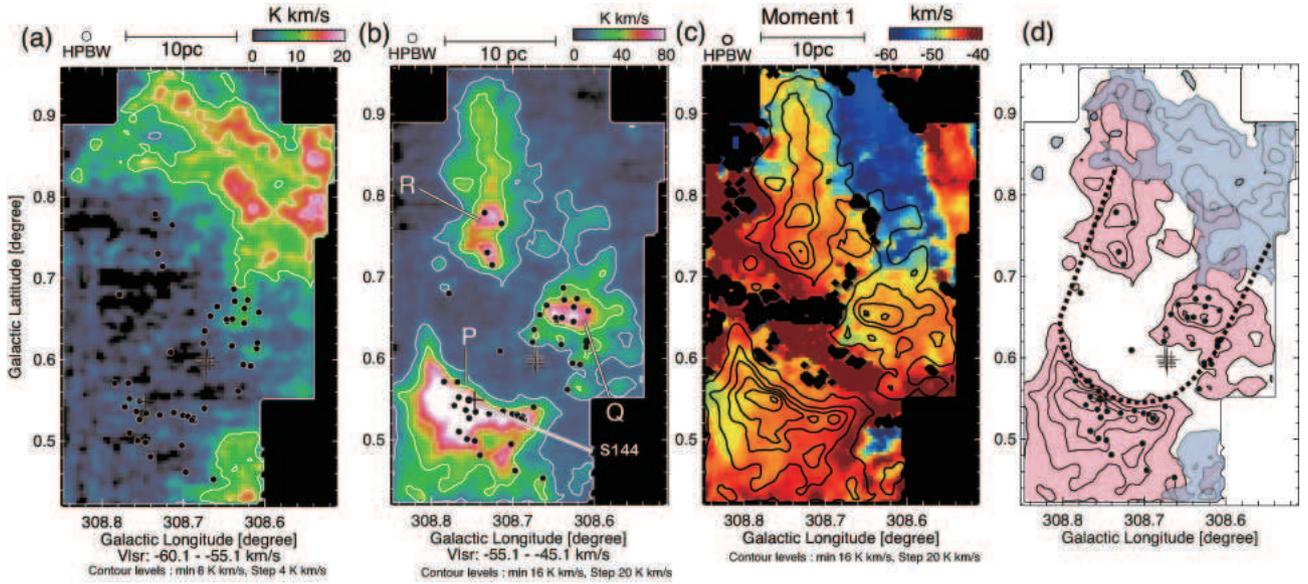}
\end{center}
\caption{
(a) The integrated intensity map of the $-57$ km s$^{-1}$ component. Black circles show YSOs identified by Liu et al (2017). Crosses depicts exciting O stars (Martins et al. 2010). 
(b) The integrated intensity map of the $-47$ km s$^{-1}$ component.  
(c) The 1st moment map of the $^{12}$CO $J=$1--0 emission, which was created for a velocity range of $-70$-- $-30$ km s$^{-1}$ using the volume voxels with the intensities of higher than 3 K.
(d) The distribution of the two clouds at $-57$ km s$^{-1}$ and $-47$ km s$^{-1}$. The dashed line indicates the projected outer boundary of the cylindrical cavity created by the collision in the $-47$ km s$^{-1}$ cloud. The width of the cavity matches approximately the size of the $-57$ km s$^{-1}$ cloud on the top of the cavity. The round shape of the cavity in the bottom is drawn so as to to fit the inner shape of the cloud P.
%(a) The schematic image of two colliding clouds in RCW 79. The blue and red cloud show $-57$ km s$^{-1}$ and $-47$ km s$^{-1}$ component, respectively. The arrow indicates the progressing direction of a collision. 
}\label{.....}
\end{figure*}

\begin{figure*}[h]
\begin{center}  \includegraphics[width=12cm]{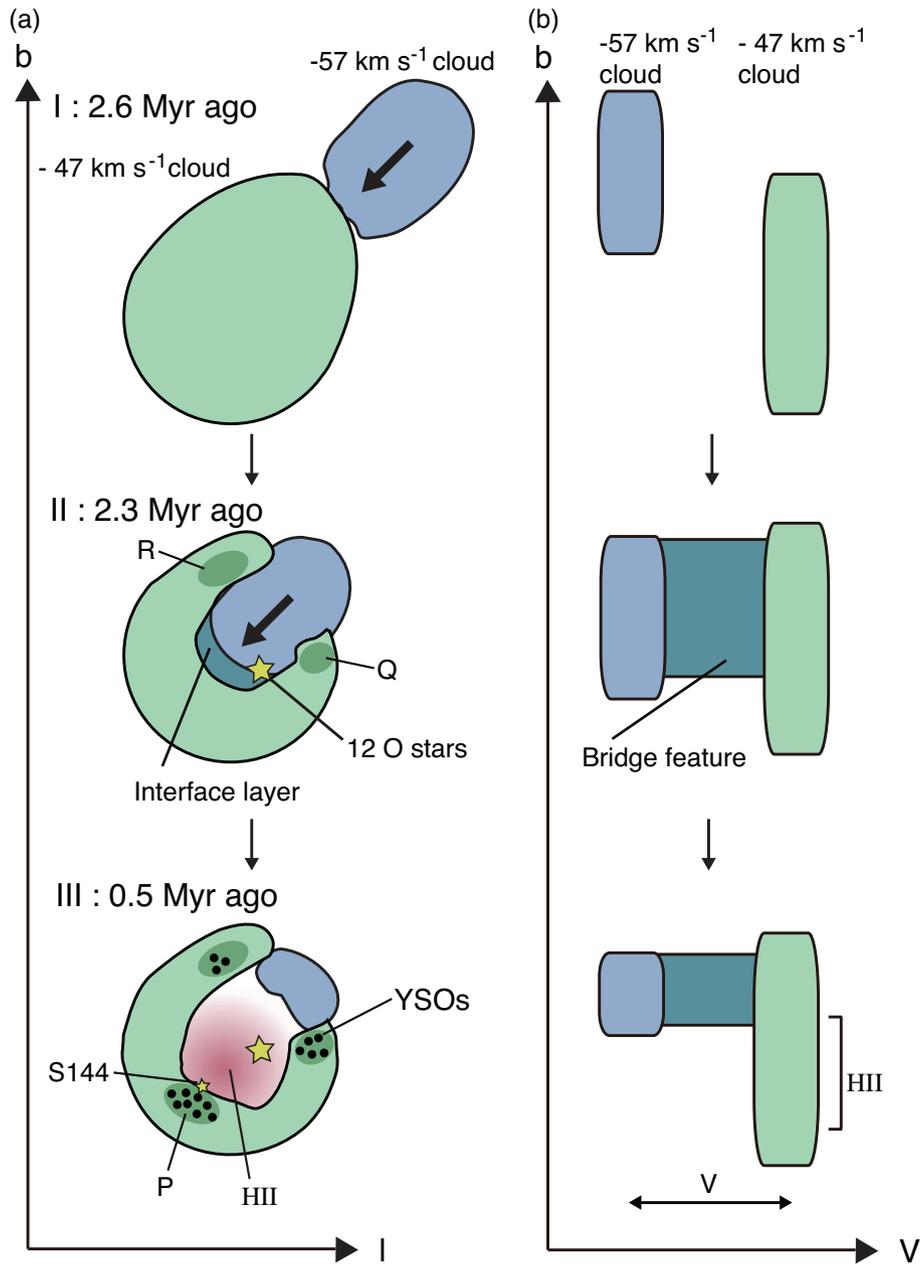}
\end{center}
\caption{ 
a) Schematic diagrams of the collision between the two clouds seen in the sky.
Stage I: The two clouds begin collision 2.6 Myrs ago.
Stage II: The small cloud ($-57$ km s$^{-1}$ cloud) created a cavity in the large cloud ($-47$ km s$^{-1}$ cloud) and the 12 O stars were formed 2.3 Myrs ago. R and Q were compressed.
Stage III: 0.5 Myrs ago the cavity was fully ionized and the small cloud inside the cavity was fully dissipated by collision and ionization. The remnant of the small cloud lies toward the top of the cavity. In P, R and Q mainly low mass stars were formed by collisional triggering, with an exception of S144 accompanying an O star.
b) Schematics of the three stages I, II, and III in the position velocity diagrams. The two clouds and the bridge feature are indicated. No deceleration by the collisional process is taken into account.
}\label{.....}
\end{figure*}

\clearpage
\appendix
\section*{CO velocity channel maps of  Mopra}
We show the velocity channel maps of the $^{12}$CO $J=$1--0 emission in Figure 8. The velocity range is between $-64.9$ and $-33.3$ km s$^{-1}$.
\begin{figure*}[h]
\begin{center}  \includegraphics[width=14cm]{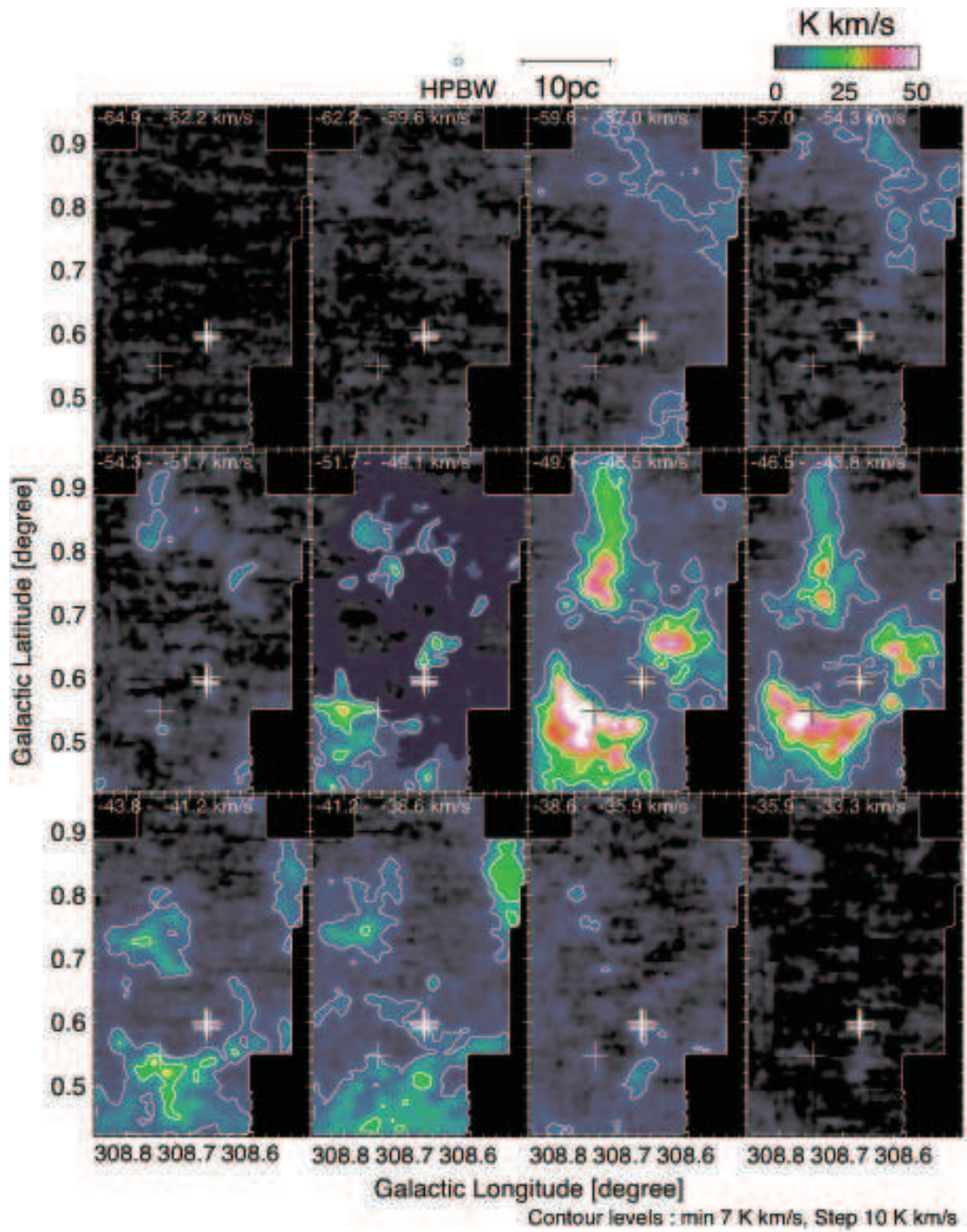}
\end{center}
\caption{Mopra $^{12}$CO $J=$1--0 velocity channel distributions toward RCW 79. Crosses depicts exciting O stars (Martins et al. 2010).}\label{.....}
\end{figure*}

%\begin{table}[h!] 
%\begin{center}
%\tbl{Model parameters (Inoue \& Fukui 2013) }{% 
%\begin{tabular}{@{}c|ccccc@{}} \noalign{\vskip3pt}
%\hline\hline
%Parameter  & note    \\ \hline
%$<n>_0$ [cm$^{-3}$] & 300 &  \\  
%$\Delta n/<n>_0$ & 0.33 &  \\ 
%$B_0$ [$\mu$G] & 20 &  \\ 
%$v_{\rm coll}$ [km s$^{-1}$] &  10 & \\ 
%Gravity & Yes & \\ 
%Resolution [pc] &  8.0/512 &  \\ 
%$M_{\rm core, tot}$ [$M_{\odot}$] & 167 &  \\ 
%$M_{\rm mag}$ [$M_{\odot}$] & 124 &  \\ 
%$M_{\rm turb}$ [$M_{\odot}$] & 41 &  \\ 
%$<n>_{\rm core}$ [cm $^{-3}$] & 8.3e4&    \\ 
%$<B>_{\rm core}$ [$\mu$G] & 2.8e2 &  \\ 
%$\Delta v_{\rm core}$  [km s$^{-1}$] & 0.97 &  \\ 
%$t_{\rm form}$ [Myr] & 0.63 &    \\ 
%\hline\noalign{\vskip3pt} 
%\end{tabular}}
%\label{tab:first} 
%\begin{tabnote}
%{\hbox to 0pt{\parbox{78mm}{\footnotesize
%\par\noindent
%\footnotemark  \par\noindent
%\footnotemark
%\phantom{0}
%\par
%\hangindent6pt\noindent
%\hbox to6pt{\,\footnotemark \hss}\unskip%
%}\hss}}
%\end{tabnote} 
%\end{center} 
%\end{table}

\end{document}